\documentclass[journal=jacsat,manuscript=article]{achemso}

\usepackage{multirow}
\usepackage[version=3]{mhchem} % Formula subscripts using \ce{}

\author{Chen Sun, Weijie Xu, Yongqi Tan, Yuqing Zhang, Zengqi Yue, Sahar Shabbir, Mengting Wu, Long Zou, Fengye Chen}
\affiliation[Shanghai Jiao Tong University]
{School of Physics and Astronomy, Shanghai Jiao Tong University, \\Shanghai 200240, P. R. China}
\author{Jin Yu}
\email{jin.yu@sjtu.edu.cn}

\title[An \textsf{achemso} demo]
  {From Machine Learning to Transfer Learning in Laser-Induced Breakdown Spectroscopy: the Case of Rock Analysis for Mars Exploration}

\abbreviations{IR,NMR,UV}

\begin{document}


\begin{abstract}
With the ChemCam instrument, laser-induced breakdown spectroscopy (LIBS) has successively contributed to Mars exploration by determining elemental compositions of the soil, crust and rocks. Two new lunched missions, Chinese Tianwen 1 and American Perseverance, will further increase the number of LIBS instruments on Mars after the planned landings in spring 2021. Such unprecedented situation requires a reinforced research effort on the methods of LIBS spectral data treatment. Although the matrix effects correspond to a general issue in LIBS, they become accentuated in the case of rock analysis for Mars exploration, because of the large variation of rock composition leading to the chemical matrix effect, and the difference in morphology between laboratory standard samples (in pressed pellet, glass or ceramics) used to establish calibration models and natural rocks encountered on Mars, leading to the physical matric effect. The chemical matrix effect has been tackled in the ChemCam project with large sets of laboratory standard samples offering a good representation of various compositions of Mars rocks. The present work deals with the physical matrix effect which is still expecting a satisfactory solution. The approach consists in introducing transfer learning in LIBS data treatment. For the specific case of total alkali-silica (TAS) classification of natural rocks, the results show a significant improvement of the prediction capacity of pellet sample-based models when trained together with suitable information from rocks in a procedure of transfer learning. The correct classification rate of rocks increases from 33.3\% with a machine learning model to 83.3\% with a transfer learning model. 
\end{abstract}

\section{Introduction}
It is generally considered that the matrix effects, both the chemical matrix effect\cite{1} and the physical matrix effect,\cite{2} represent a critical issue in analyses with laser-induced breakdown spectroscopy (LIBS) for either qualitative classification or quantitative determination.\cite{3} Suitable solutions with respect to such consideration become primordially determinant for applications as important as LIBS analysis of rocks in Mars explorations,\cite{4} where the targeted scientific goals, searching for the present and past water activities and the traces of the life, as well as the study of the Mars habitability,\cite{5,6,7} rely, at least partially, on the reliability and the exactitude of the analytical data that one can extract from LIBS spectra recorded by LIBS instruments embarked on Mars rovers.\cite{8} Certainly the diversity of chemical composition of Mars rocks has been studied by the precedent missions, the absence of real samples from Mars, except meteors, requires a large number of laboratory rock standard samples to be prepared with Earth natural rocks or by mixing pure chemical compounds, in order to cover the chemical variety of Mars rocks. It was the purpose of the sets of laboratory standard rock samples prepared and used by the ChemCam team for training and validation of Mars LIBS spectral data processing models. The number of involved samples was first 69,\cite{9} they were further increased to 408 in order to offer a larger covering of the chemical and mineral compositions of Mars rocks.\cite{10} It is useful and important to point out that all the laboratory rock standards were prepared in the form of pressed powder disks, glasses, and ceramics to minimize heterogeneity in the scale of LIBS observation of typically several hundred µm. Such sample preparation leads to obvious differences in surface morphological and physical properties between laboratory standards and real rocks analyzed by LIBS on Mars, differences from which physical matrix effect can result. With this concern, the effect of surface roughness on the hydrogen emission line has been investigated.\cite{11} Our recently published work\cite{12} observed and analyzed the performance of a machine learning-based model\cite{13} trained with a set of pressed rock powder pellets for total alkali-silica (TAS) classification\cite{14} of rocks in their natural state. Important degradation of the model prediction performance compared to the prediction for pellet samples has been observed. Such degradation prevents the models trained with laboratory standards from reliable predictions with LIBS spectra acquired in situ on real rock samples, a situation which can become crucial in an application as we mentioned above, of in situ LIBS analysis of Mars rocks, since we are not yet able to bring back samples from Mars. 

In order to search a solution for the discussed issue, in this work, transfer learning has been introduced for LIBS spectral data treatment. Transfer learning is considered in machine learning when the knowledge gained while solving one problem is required to be applied to a different but related problem.\cite{15} Its necessity comes from the fact that a major assumption in many machine learning and data mining algorithms is that the training and the model-targeted data should share the same feature space and have the same distribution.\cite{16} It is unfortunately not the case for the application scenario that we consider. Moreover, transfer learning has recently emerged as a new learning framework to address the problem of insufficient training data in an application (target domain) with the help of the knowledge learnt from a related application with the facility to get sufficient training data (source domain).\cite{17} Such strategy fits the requirement of LIBS analysis of Mars rocks, where sufficient laboratory standards can be prepared as the source domain, while real Mars rock samples are not yet available as the target domain. According to the specific contents in the “knowledge” to be transferred, we can distinguish feature-representation-transfer, where parts of relevant features respectively from the both source and target domains are further selected for their low sensitivity to the difference between the two domains, to form a common set of transfer features contributing to the training of a transfer learning model.\cite{18} Instance-transfer is another specification of transfer learning where data from the both source and target domains participate in the transfer learning model training, with a conditional testing of the relevance of each data from the source domain for its effect in improving the performance of the model during a cross-validation with data from the target domain.\cite{18} A weight is then applied to each source domain data according to its efficiency in improving the performance of the model for predicting with target domain data.

More specifically, in our experiment, on the basis of the LIBS spectra from a set of laboratory standard samples in the form of pressed pellets, a machine learning-based multivariate model was trained and used to predict the concentrations of major oxides, $SiO_2$, $Na_2O$ and $K_2O$, with the spectra from the corresponding natural rock samples, concentrations necessary for the TAS classification of the rocks. The influence of the physical matrix effect was observed. A modified model training procedure was then applied with the adjunction in the training sample set, of a small number of natural rocks with certified concentrations of the major oxides. A transfer learning-based model was thus trained with the implementation of feature-representation-transfer and instance-transfer, and used to predict the concentrations of the major oxides for the rock samples. In the following, we will first briefly present the used samples, the experimental setup and protocol. We describe then the methods of machine learning- and transfer learning-based model trainings, before the performances of the both models being presented and compared in order to draw the conclusion of the work.

\section{Samples and Experimental setup}

\subsection{Sample}

\begin{figure}
\centering\includegraphics[width=10cm]{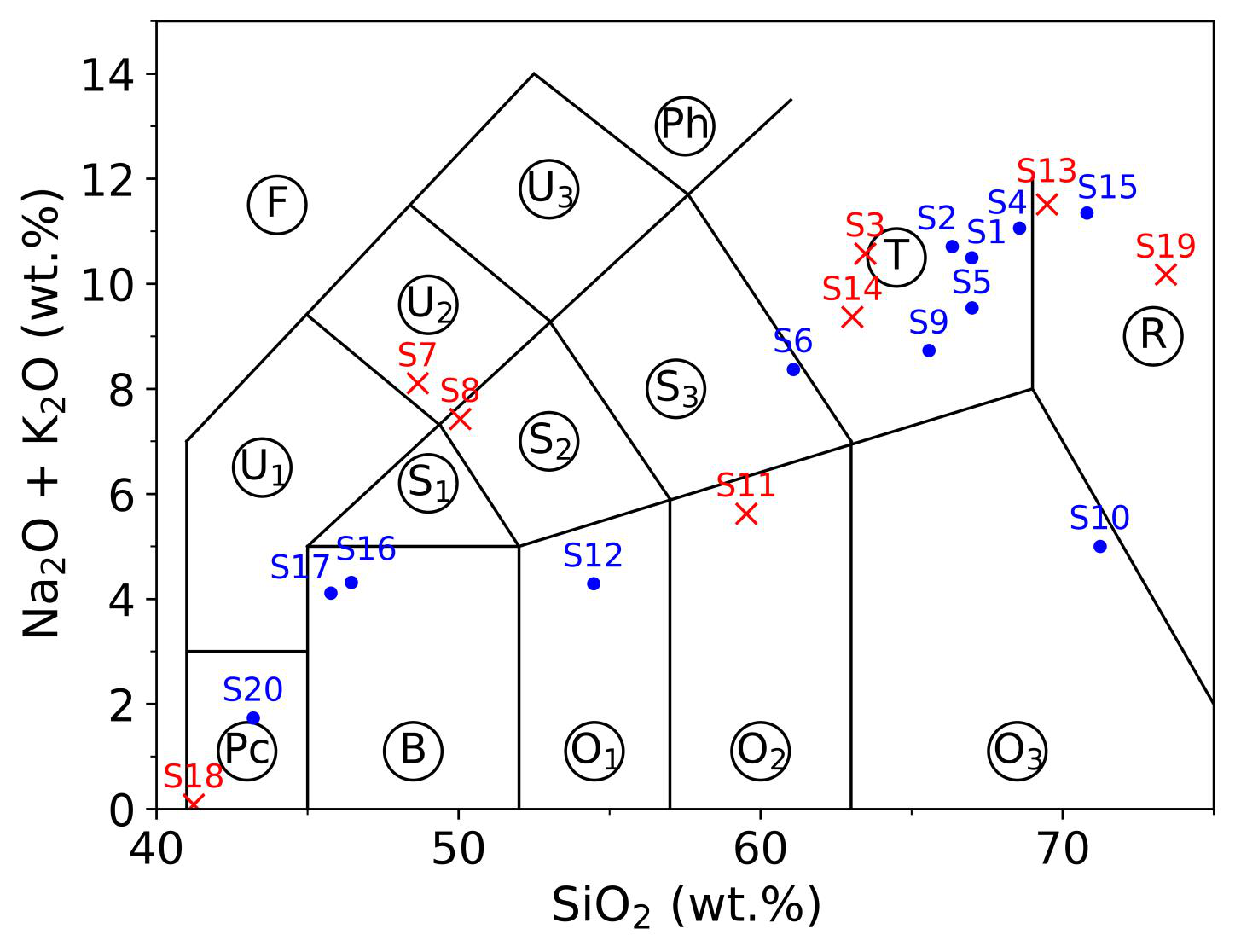}
\caption{Presentation of the used rock samples in a TAS diagram according to their major oxide concentrations determined using XRF. The short notations of the 15 fields (surrounded by circle) are according to Reference 14. Eight rock were selected as training samples (represented by red crosses in the figure): S3, S7, S8, S11, S13, S14, S18 and S19. The rest of the 12 samples were used as validation samples (represented by blue dots in the figure).}
\label{fig:fig1}
\end{figure}

In this work, 20 natural terrestrial rocks were used as samples for LIBS analysis. The rocks were first washed using alcohol and distilled water before any further preparation. All the rocks were analyzed in 2 different forms. Raw rocks: LIBS measurements took place on the natural surface of each rock. Pellets: a part of each rock was crushed and ground into powder by a laboratory mill and then sieved by a 300-mesh screen (grain size <50 $\mu m$). A binder (microcrystalline cellulose powder) with similar particle size was mixed with the rock powder at a weight ratio of 20\%. One gram of the obtained powder was pressed under a pressure of 850 MPa for 30 minutes to form a pellet of 15 mm diameter and ~ 2 mm thickness. The composition, with especially the concentrations of major oxides, $SiO_2$, $Na_2O$, $K_2O$, of each rock was determined by XRF with the pellets, which allowed presenting the rocks in a TAS diagram as shown in Figure \ref{fig:fig1}.

\subsection{Experimental setup}

\begin{figure}
\centering\includegraphics[width=10cm]{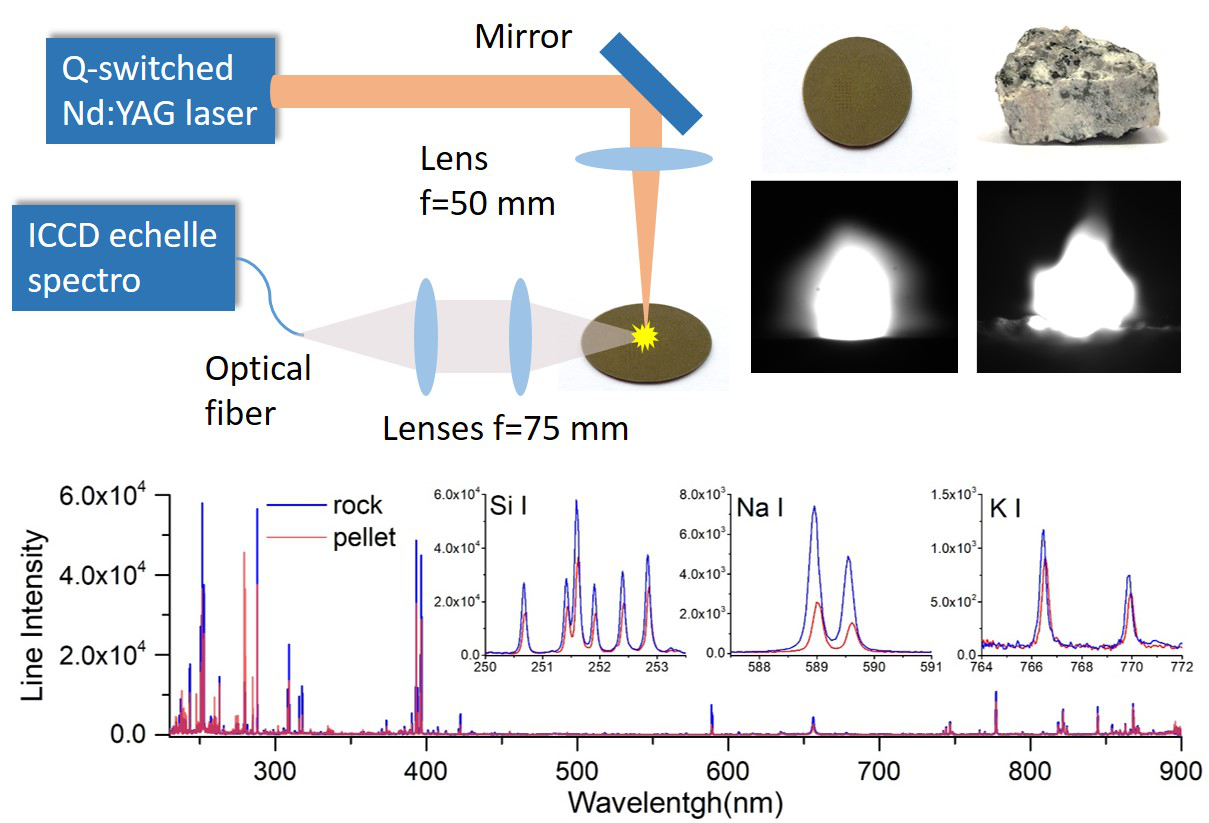}
\caption{Schematic presentation of the used experimental setup, together with plasma images respectively induced from a pellet and a rock, and typical LIBS spectra showing difference in emission intensity of Si, Na and K between a pellet and the corresponding rock sample.}
\label{fig:fig2}
\end{figure}

A detailed description of the used experimental setup can be found elsewhere.12\cite{ACS2007} Briefly as shown in Figure 2, a Q-switched Nd:YAG laser operated at a wavelength of 1064 nm with a pulse duration of 7 ns and a repetition of rate of 10 Hz, was used to ablate the samples with a pulse energy of 8 mJ. A lens of 50 mm focal length focused laser pulses about 0.86 mm below the surface of a sample. The diameter of the focused laser spot on the sample surface was estimated to 150 µm, corresponding to a laser fluence on the sample surface of about 45 J/cm2, or an irradiance of about 6.5 GW/cm2. Emission from the generated plasma was collected by a combination of two quartz lenses with a same focal length of 75 mm into an optical fiber of 50 µm core diameter. The output of the fiber was connected to the entrance of an echelle spectrometer equipped with a ICCD camera (Mechelle 5000 and iStar, Andor Technology) which provided a wide spectral range from 230 nm to 900 nm with spectral resolution power of 5000. The ICCD camera was triggered by laser pulses with a delay and a gate width of respectively 500 ns and 2000 ns. A lateral CCD camera (not shown in the figure) allowed capturing time-integrated plasma images as shown in the inset of Figure 2. The samples were mounted on a 3D translation stage allowing recording replicate spectra on a sample surface with an ablation crater matrix, while keeping a constant distance between the focusing lens and the sample surface.

\subsection{Spectrum acquisition.}

For each sample, 50 replicate spectra were taken on 50 different ablation craters, and each crater received 25 successive laser shots. The emission spectra induced by the first 5 laser shots were removed in order to avoid surface contaminations, and those induced by the subsequent 20 laser shots were accumulated to produce a replicate spectrum. Such procedure also intended to reduce the difference in surface roughness between the 2 different forms of the samples. In total, 2000 spectra were recorded for the 20 rocks with the 2 different forms for each of the rocks. Typical spectra are presented in the inset of Figure 2 for a pellet and the corresponding rock sample, showing different line intensities of Si, Na and K. Such difference corresponds well to that observed for plasma images induced with the different types of samples. 

\section{Data Treatment method.}

\begin{figure}
\centering\includegraphics[width=10cm]{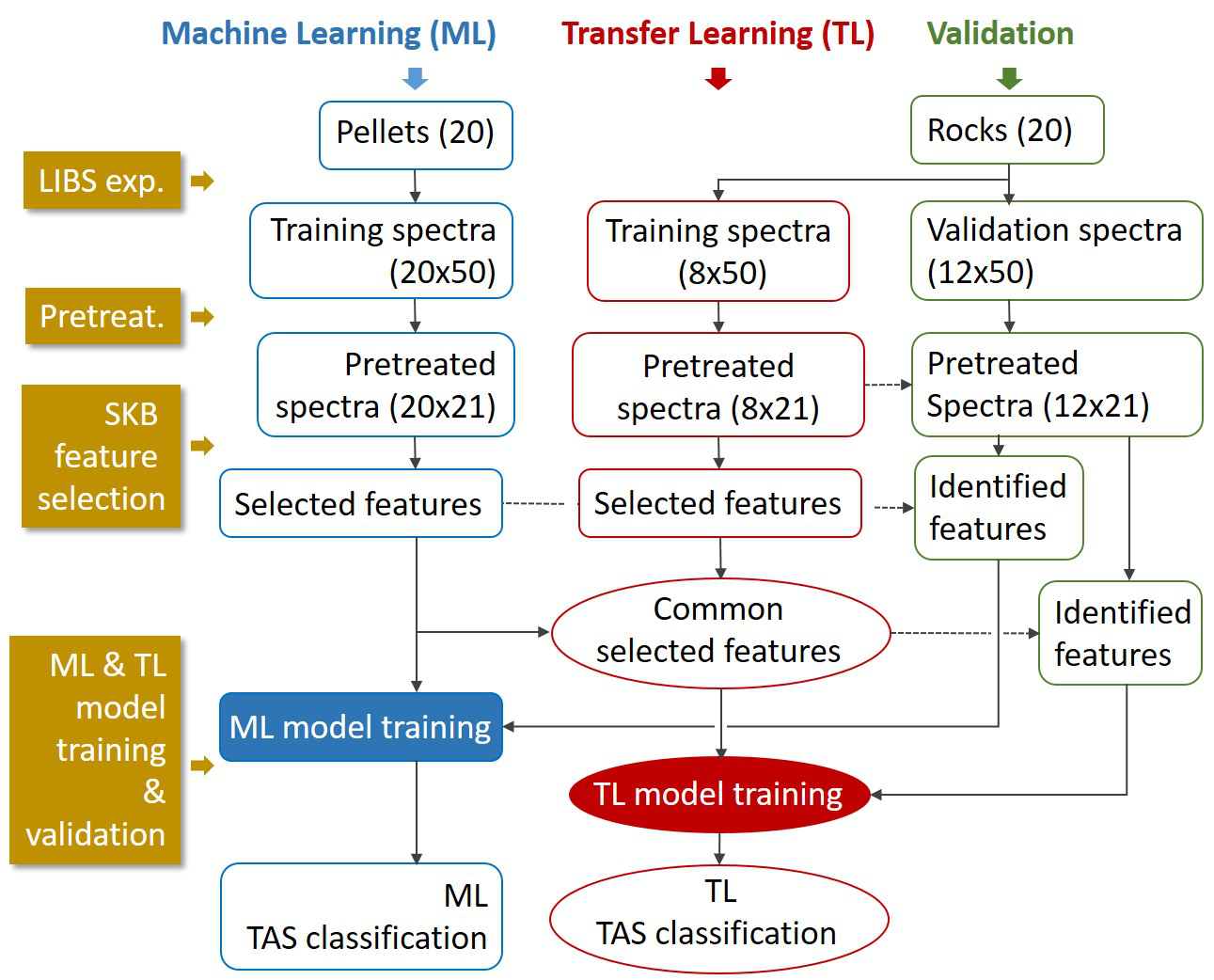}
\caption{General flowchart used in this work allowing a comparative study between the performances of a machine learning (ML) model and those of a transfer learning (TL) model.}
\label{fig:fig3}
\end{figure}

The general data treatment flowchart used in this work is shown in Figure 3. Several steps can be distinguished: pretreatment, feature selection, machine learning (ML) and transfer learning (TL) model trainings, and model validation. The LIBS spectra from the pellet samples were used as training data, while those from the rocks were separated into a training data set containing 8 samples and a validation data set containing 12 samples.

\subsection{Data treatment}

The pretreatment consisted in the following operations. i) Average in order to reduce experimental fluctuations and the effect of sample inhomogeneity: For each sample, the 50 raw spectra were averaged in a procedure where an averaged spectrum was calculated with a first group of randomly selected 30 spectra. The rest 20 spectra then replaced one by one, a spectrum in the first group, each time the new group of 30 spectra was averaged to generate 20 other average spectra. 21 average spectra were generated for each sample. ii) Baseline correction: an average spectrum was decomposed into a set of cubic spline of undecimated wavelet scales, the local minima were found, then the spline function was interpolated through the different minima to construct the spectral baseline which was removed.\cite{19} iii) Normalization: Baseline-corrected average spectra were normalized with their respective total intensity (the area under the spectrum). iv) Standardization: Standard normal variate (SNV) transformation was respectively applied to the normalized baseline-corrected average spectra of the pellets (20×21=420 spectra) and the training set of the rock samples (8×21=168 spectra). Within a given sample set, for each channel in a spectrum (22161 channels in total), the variation range of the intensity value over all the samples was transformed into a range with a mean value equal to 0 and a standard derivation ($SD$) equal to 1. The parameters determined in the standardization of the training set of the rock samples (the mean and the ) were applied to the validation set of the rock samples (12×21=252 spectra) by assuming a same statistical distribution of the data for all the rock samples. The ensemble of the above pretreatments generalized pretreated spectra. 

\subsection{Spectral feature selection.}

SelectKBest algorithm\cite{20} was respectively applied to the pretreated spectra of the pellets and the training set of the natural rocks, and successively for the 3 concerned oxides. Within a sample set, for each spectral channel, covariance was calculated between the channel intensity and the concentration of the concerned compound in the corresponding sample, over all the spectra of the sample set. A score was then calculated as a function of the covariance according to the definition given in Reference 13. Different scores, $\rho_{i,j}$, were thus associated to the spectral channels, with 2 index and a value varying from 1 to 22161, which ranks the them from the lowest score to the highest one. Such procedure was applied to the 2 sample sets ($i=1$: pellets, $i=2$: calibration set of rocks) and the 3 concerned oxides ($j=1$: $SiO_2$, $j=2$: $Na_2O$, $j=3$: $K_2O$). A feature selection procedure first identified 100 highest ranked spectral channels respectively for each of the 3 oxides in each of the 2 sample sets. Pearson’s correlation coefficient\cite{21} related to the above mentioned covariance was calculated for the 6 groups of 100 selected features. The results showed that all the selected features had a Pearson’s coefficient larger than 0.75.

As we can see in the Figure 3, the 3 groups of 100 features selected for the 3 oxides in the pellet sample set were directly used to respectively train the calibration models for the 3 oxides base on a back-propagation neural network (BPNN). The training algorithm which involved stochastic gradient descent (SGD) and mini-batch stochastic gradient descent (MSGD) optimization iterations, as well as cross validation with randomly generated statistical equivalent data configurations, has been presented in detail in Reference 13. We will not in this paper go into more detail about such training algorithm.  

For the transfer learning model training, and according to the principle of feature-representation-transfer discussed above, an ensemble of common selected features was identified between the pellet and the training rock sample sets, by calculating a total ranking index $\rho_j = \rho_{1,j}+\rho_{2,j}$. A feature selection procedure then retained the 100 highest ranked features according to the value of $\rho_j$ from the highest one to the lowest one, respectively for the 3 oxides. These groups of features were fed into the transfer learning model training algorithm. As an example, the results of feature selection for $Na_2O$ are shown in Figure 4, although similar behaviors can also be observed in the feature selections for the other 2 oxides.

\begin{figure}
\centering\includegraphics[width=10cm]{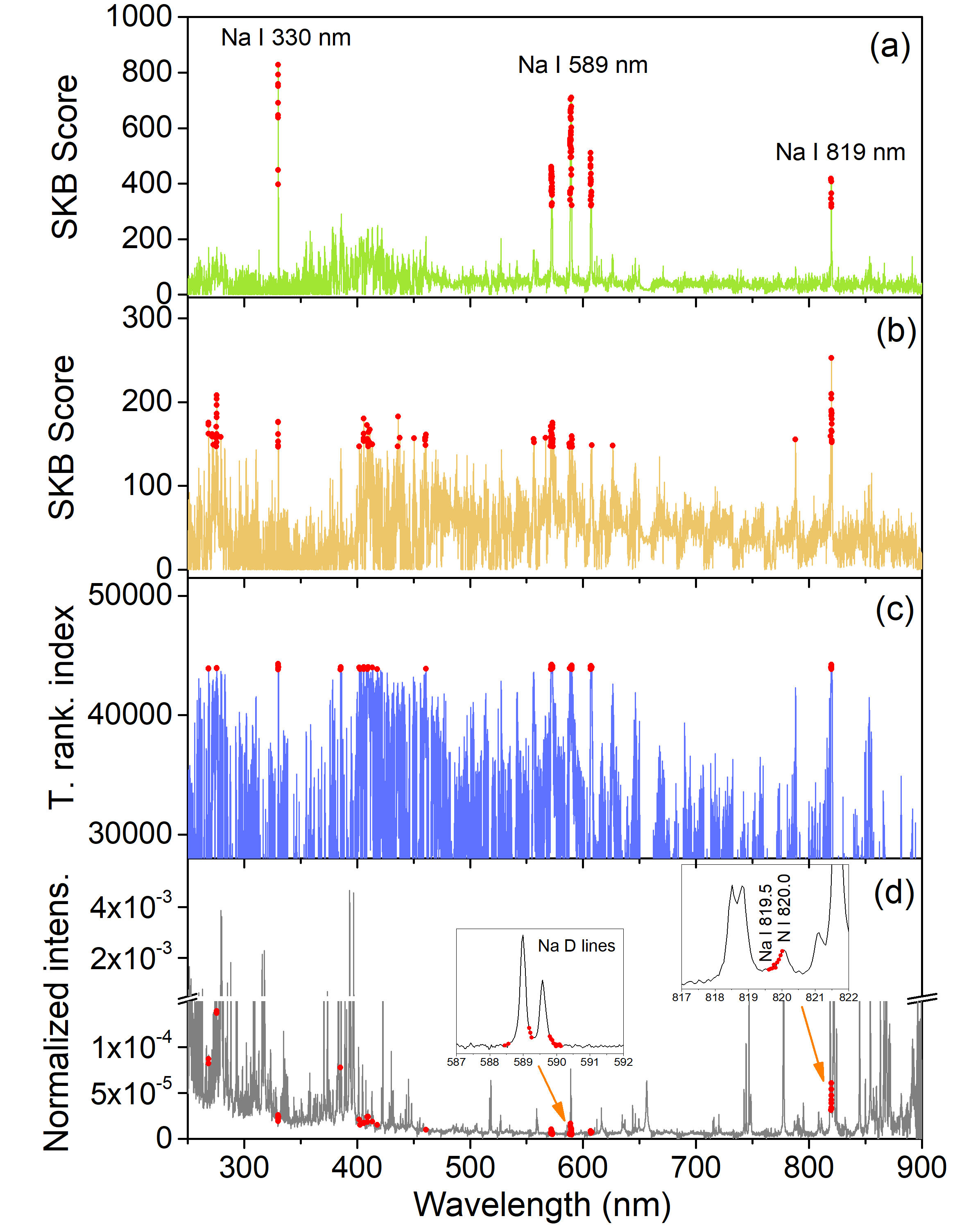}
\caption{Results of feature selection for $Na_2O$: (a) for pellets and (b) for calibration rocks, SKB scores of all the spectral channels with in red dots the 100 selected features; (c) total ranking index  of all the spectral channels, together with in red dots the 100 common selected features; (d) a typical normalized average spectrum from a pellet sample, together with in red dots the 100 common selected features, with 2 insets showing enlarged parts of the spectrum around 589 nm and 820 nm.}
\label{fig:fig4}
\end{figure}

In Figure 4 (a), we can see that for the pellets, the spectral channels with high SKB scores are clearly concentrated around several Na emission lines: Na I 330.24 nm and 330.30 nm lines, the sodium D lines: Na I 588.99 nm and 589.59 nm lines (2 groups of ghost lines around 572.1 nm and 606.9 nm are recorded due to these strong lines), Na I 818.33 nm and 819.48 nm lines. For the calibration rocks in Figure 4 (b), the selected features are distributed among other channels with a significant decrease of the scores for all the important features. This means that the physical matrix effect perturbs the inherent correlation between the emission line intensities of an element and its concentration in the material, and reduces therefore their importance in the concentration determination. In the same time, other spectral channels, such as those around 275 nm and between 410 nm and 460 nm, get relatively higher scores. This means that they become important in the determination of elemental concentration when using a model based on the calibration set of the rock samples. These features, representative of the rock samples, are thus included in the common selected features for transfer learning model training. Figure 4 (c) shows the total ranking index for $Na_2O$,  and in red dots, the 100 common selected features. These features are indicated in a typical spectrum in Figure 4 (d) in red dots. We can see that, beside the features related to the Na emission lines, some features important for the rock samples are included. A more detailed peak identification using the NIST database,\cite{22} shows the contributions from Fe II 268.475 nm and 275.57 nm lines, Si II 385.366 nm and 385.602 nm lines, and the probable contributions from K I 404.414 nm and 404.721 nm lines, Ca I 409.85 nm lines, and Si II 412.807 nm and 413.089 nm lines. A selected feature around 461 nm cannot have easy interpretation.

In the insets of Figure 4 (d), 2 parts of the spectrum are enlarged. The inset around 589 nm shows the sodium D lines together with the selected features in red dots. We can see that the selected features are located in the side parts of a line profile, while the central part of the line is not retained by the feature selection procedure. This is due to self-absorption of the strong resonant Na D lines, which affects much more the central part of the spectral lines. This observation shows the capability of the feature selection procedure to reduce the influence of self-absorption. The second inset in Figure 4 shows an enlarged part of the spectrum around 820 nm, where we can see the selected features related to the Na I 819.5 nm line in red dots. Due to the interference with the N I 820.0 nm, only the short wavelength part of the spectral profile around 820 nm is included in the selected features, showing the efficiency of the selection procedure to avoid the influence of spectral interference. 

\subsection{Transfer learning-based calibration model training.}

A training algorithm of the transfer learning model was developed in this work on the basis of that used for machine learning model training presented in detail in our previous publication\cite{13} and used in various application scenarios.\cite{12,23,24,25,26} The flowchart of transfer learning model training is shown in Figure 5. We can distinguish 3 main steps: data formatting; model training by optimization through iteration loops and model validation. Training was respectively performed for the 3 concerned oxides, resulting in 3 specific models. 

The optimization and the assessment of the models were performed in this work using a certain number of indicators specified bellow: determination coefficient of a linear regression $r^2$ indicating the correlation of the calibration data with respect to the regression model, limit of detection $LOD$ of a model, average relative error of calibration $REC(\%)$ assessing the accuracy of a calibration model to be tested, average relative error of test $RET(\%)$ assessing a tested model to be validated, average relative error of prediction $REP(\%)$ assessing the trueness of the model-predicted concentrations, average relative standard deviation $RSD(\%)$ assessing the precision of the model-predicted concentrations. The mathematical definitions of these parameters can be found elsewhere, in particular in References 13 and 27.

\begin{figure}
\centering\includegraphics[width=7cm]{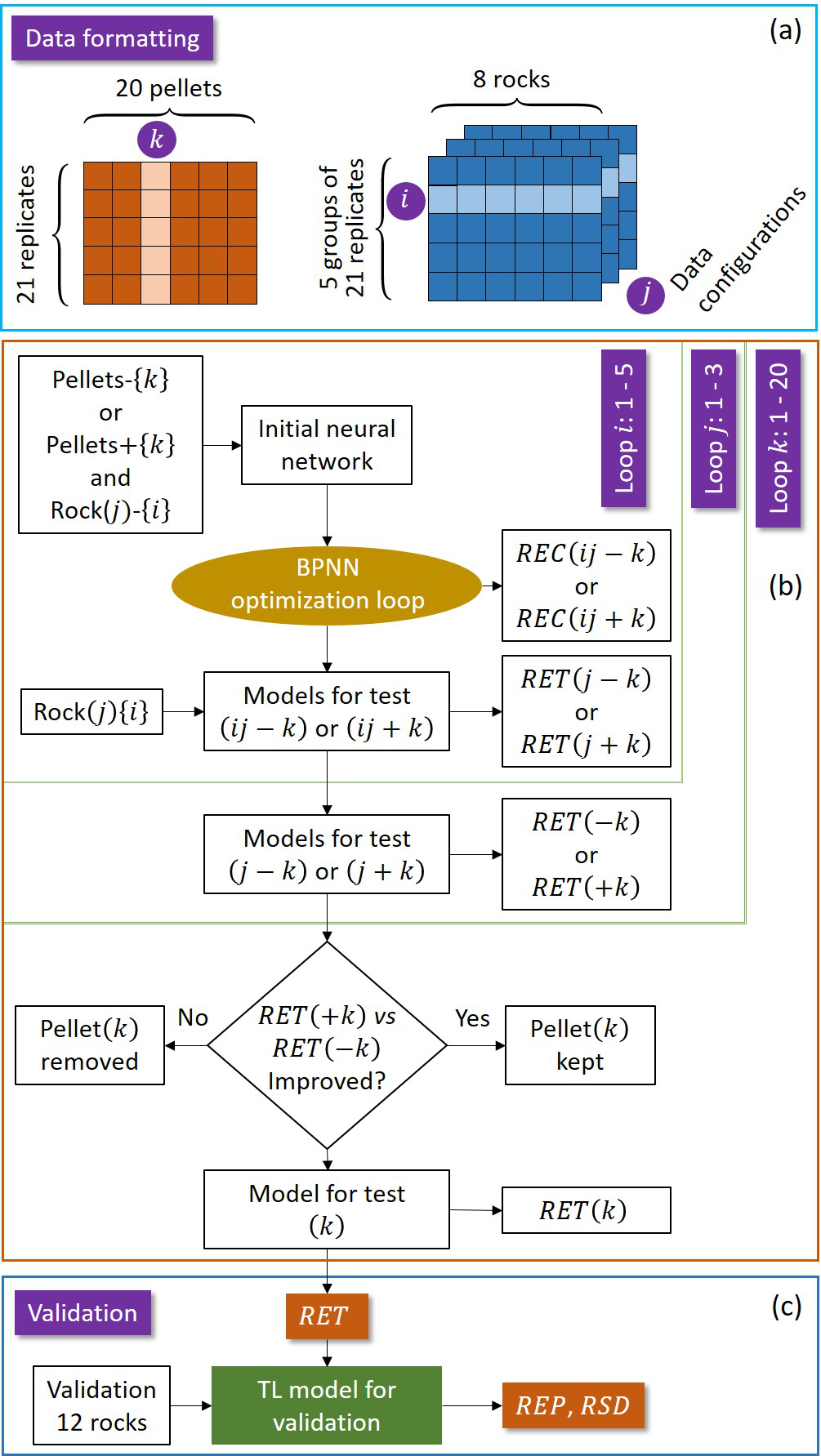}
\caption{Flowchart of the transfer learning model training with the implementation of feature-representation-transfer and instance-transfer.}
\label{fig:fig5}
\end{figure}

\subsection{Data formatting.}
According to the above discussed principles of feature-representation-transfer and instance transfer in transfer learning, spectra from the pellet samples (the source domain) and those from the calibration set of the rock samples (the target domain), with respectively 100 common selected features, participated in the training process. All the 20 pellet samples with their average replicate spectra were initially involved in the training data set. These spectra were organized in a given data configuration where the replicate spectra for each sample were arranged in an arbitrary order. The efficiency of each pellet was tested within an iteration loop where the $RET$s with and without the spectra from the pellet were compared in order to decide the exclusion or the definitive inclusion of the pellet in the transfer learning model training sample set. It was why the ensemble of replicate spectra associated to one of the 20 pellets was indexed with  which went from 1 to 20 (Figure 5 (a)) for the test the efficiency of all the pellet samples during the training process. Eight chosen rock samples contributed to the transfer learning model training data set (S3, S7, S8, S11, S13, S14, S18 and S19 in Figure 1). In particular, they were used in a cross validation process during the optimization of the neural network. It was why that the spectra of this data set were first organized in different data configurations where each configuration $j$ corresponded to a certain arrangement of replicate spectra for each rock (Figure 5 (a)). Data configurations were all statistical equivalent since the order of a replicate spectrum of a sample was a dummy index. The number of different data configurations were limited to 3 in this work because more configurations did not bring further improvement of the model as tested in the work. For a given configuration $j$, the spectra were further organized into 5 groups of replicates containing respectively 4, 4, 4, 4 and 5 spectra, respectively. A new index $i$ was introduced to designate a group of replicate spectra of all the rock calibration samples as shown in Figure 5 (a). During the model training process, the index $i$ went from 1 to 5 within an iteration loop of cross validation, indicating each time the validation group of replicates. 

\subsection{Model training by optimization.}

A 3-layer back propagation neural network (BPNN) similar to that used in Reference 13 was employed in this work for the transfer learning model. The network was composed by an input layer of 100 neurons corresponding to the 100 common selected features of each input spectrum; a hidden layer 5 neurons and an output layer with a single neuron corresponding to the targeted compound concentration. The function of the network was therefore to map an input spectrum (a vector of 100 dimensions) to a scalar which can be considered as the module of a vector in a hyperspace. The precision of the mapping was improved during the training process through different iteration loops under the supervision of the targeted concentrations using the model performance indication parameters specified above. 

As shown in Figure 5 (b), 3 hierarchized iteration loops ($i,j,k$, among them $i,j$ are doubled loops:  for $\pm k$) surrounding the BPNN optimization loop, performed the supervised optimization of the model: 

-a doubled inner loop for $i=1$ to 5: for the double cases of a given sample $k$ in the pellet sample set being excluded ($-k$) and included ($+k$) in the training data set, and a given data configuration  of the rock spectra, the network was optimized within a cross-validation process where the model was trained using the ensemble of 4 groups of replicate spectra, of for example $i = 2,3,4,5$ with respectively 4, 4, 4 and 5 spectra. The resulted $REC(ij-k)$ and $REC(ij+k)$ were calculated for the respectively optimized Models for test $(ij-k)$ and $(ik+k)$. These models were then tested using the ensemble of replicate spectra of the rest $i^{th}$ group, $i=1$ for example, generating $RET(j-k)$ and $RET(j+k)$, together with the optimized Models for test $(j-k)$ and $(j+k)$.

-A doubled intermediary loop for $j=1$ to 3: in this loop, the above discussed loop $i$ was executed with 3 independent rock data configurations for the 2 cases of a given sample $k$ in the pellet set being excluded from or included in the training data set. The model was further optimized. Corresponding calculation of the values of $RET$ resulted in $RET(j-k)$ and $RET(j+k)$. 

-An outer loop for $k=1$ to 20: in this loop the above discussed loop $i$ and loop $j$ were executed for each of the 20 pellet samples successively assigned as the $k$ pellet. For a given pellet, $RET(-k)$ and $RET(+k)$ were compared. If an improvement was observed with the sample $k$, it was kept in the training sample set, otherwise it was removed. This loop generated a Model for test $(k)$ for each considered pellet sample with the corresponding $RET(k)$. The optimization process finally generated a Model for valuation with a minimized $RET$.  

\subsection{Model validation.} 

The resulted transfer learning model was validated by the pretreated spectra from the validation set of the rock samples (12 spectra) with the identified features according to the common selected features between the pellet sample set and the training set of the rock samples. The parameters assessing the performance of the model for prediction, $REP$ and $RSD$ were calculated. These parameters would indicate the performance of the model when used for predictions with LIBS spectra from unknown rock samples.

\section{RESULTS AND DISCUSSION}

\subsection{Analytical Performances with the machine learning model.}

We first present the results obtained with the models trained with the 20 pellet samples and validated with the 12 validation rock samples respectively for the 3 concerned oxides, $SiO_2$, $Na_2O$ and $K_2O$. The training method described in Reference 13 was implemented in this work to train a neural network. The training procedure was similar to the inner (loop $i$) and the intermediary (loop $j$) iteration loops used in the transfer learning model training (Figure 5 b) with a similar neural network structure. As shown in Figure 3, the input variables were the 100 selected features in a spectrum of a pellet sample for the training and the 100 identified features in a spectrum of a validation rock sample for the validation. For the cross-validation optimization in the training process, similarly as for the transfer learning model training, $3\times 5$-fold iterations were performed with 3 randomly organized pellet spectrum data configurations and 5 replicate groups of respectively 4, 4, 4, 4 and 5 spectra for each data configuration. A larger number of data configurations did not lead to a significant improvement of the model as shown by the tests in our work. The training process was executed for the 3 concerned compounds resulting in 3 prediction models respectively for $SiO_2$, $Na_2O$ and $K_2O$. The obtained results are shown in Figure 6. The extracted parameters for the assessment of the model performances are presented in Table 1. 

In Figure 6 and Table 1, We can see that the machine learning calibration models trained with pellet samples present good performances in terms of the usual assessment parameters including $r^2$, $LOD$, $REC$ and $RET$. As we have pointed out in Reference 12, this indicates an efficient chemical matrix effect correction with machine learning. At the same time, we can remark a large degradation in the performance when the model was validation by the rock validation samples, in terms of $REP$ and $RSD$. Figure 6 shows that the use of the pellet models for prediction with the spectra from the rock validation samples can lead to systematic bias, with a shift of the linear regression of the validation data with respect to the model as shown in Figure 6 (b) for $Na_2O$, as well as variance, with a change of the slope of the linear regression of the validation data with respect to the model as shown in Figure 6 (a) and (c).

These results show the effect of the physical matrix effect when the models trained with pellet samples were used for prediction for rock samples. As a consequence, the TAS classification of the validation rocks with the pellet machine learning models resulted in an unsatisfactory performance as shown in Figure 7. In this figure, the reference position in the TAS diagram of each rock determined by the compound concentrations measured using XRF (as shown in Figure 1) is indicated with a colored solid circular point. With the same color, the position predicted by the pellet machine learning models for the same rock is represented by a cross with error bars. More precisely, the cross represents the mean position calculated with the 21 pretreated validation spectra. The error bars represent the standard deviations ($\pm SD$) of the concentrations, in particular the vertical error bar was obtained by summing the $SD$ for the 2 concerned compounds. A dash-dot line further links the reference and the predicted positions of a same rock sample in order to explicitly indicate their correspondence. Such presentation thus allows calculating the rate of correct classification, $\rho$ , for the validation rock samples in a TAS diagram according to their compound concentrations determined using the machine learning calibration models, as compared to their reference positions determined by the XRF concentrations. If the pellet model-predicted position of a sample stays in the same TAS field as its XRF reference position, it is correctly classified. In figure 7, we can only see 4 correct classifications (S1, S2, S5 and S10), corresponding to a correct classification rate of 33.3\%.

\begin{figure}
\centering\includegraphics[width=7cm]{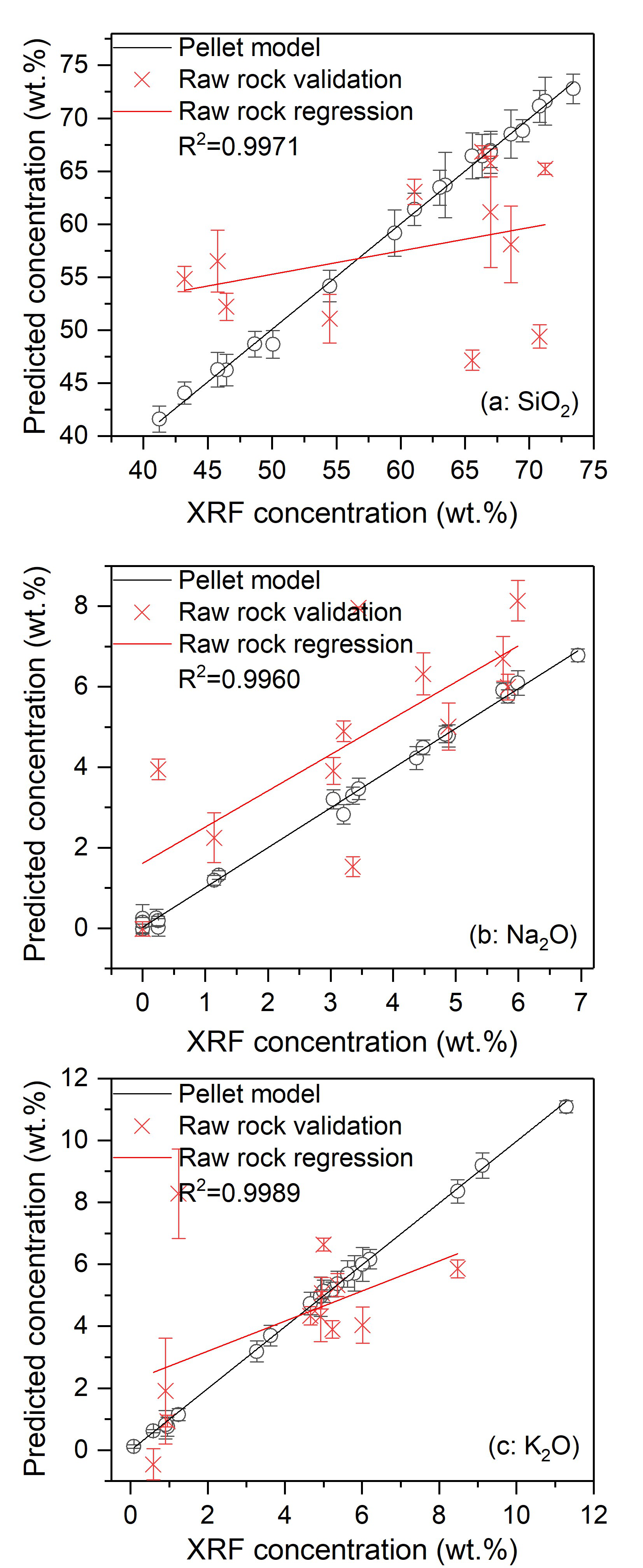}
\caption{Machine learning-based calibration models trained with the spectra from the pellet samples (black lines) together with the calibration data (black open cycles) for the 3 compounds $SiO_2$ (a), $Na_2O$ (b) and $K_2O$ (c). Validation data from the rock validation samples are presented in red crosses, their linear regressions in red lines. The error bars of the presented data correspond to the standard deviations ($\pm SD$) of the predicted concentrations over the 21 pretreated spectra for a given sample.}
\label{fig:fig6}
\end{figure}

\begin{table}
  \caption{Parameters assessing the calibration and prediction performances of the machine learning calibration models for $SiO_2$, $Na_2O$ and $K_2O$.}
  \label{tbl:1}
  \begin{tabular}{llllll}
    \hline
    \multicolumn{2}{c}{Compound} & $SiO_2$ & $Na_2O$ & $K_2O$ & Average\\
    \hline
    \multirow{5}{*}{Calibration} & $r^2$ & 0.997 & 0.996 & 0.999 & 0.997\\
    & $Slope$ & 0.994 & 0.988 & 0.997 & 0.993\\
    & $LOD(\%)$ & 5.14 & 0.622 & 1.01 & 2.26\\
    & $REC(\%)$ & 5.61 & 9.84 & 3.75 & 6.40\\
    & $RET(\%)$ & 7.42 & 10.5 & 5.20 & 7.72\\
    \hline
        \multirow{2}{*}{Validation} & $REP(\%)$ & 13.76 & 176.2 & 82.28 & 90.75\\
    & $RSD(\%)$ & 21.94 & 9.440 & 271.8 & 101.1\\
    \hline
  \end{tabular}
\end{table}

\begin{figure}
\centering\includegraphics[width=7cm]{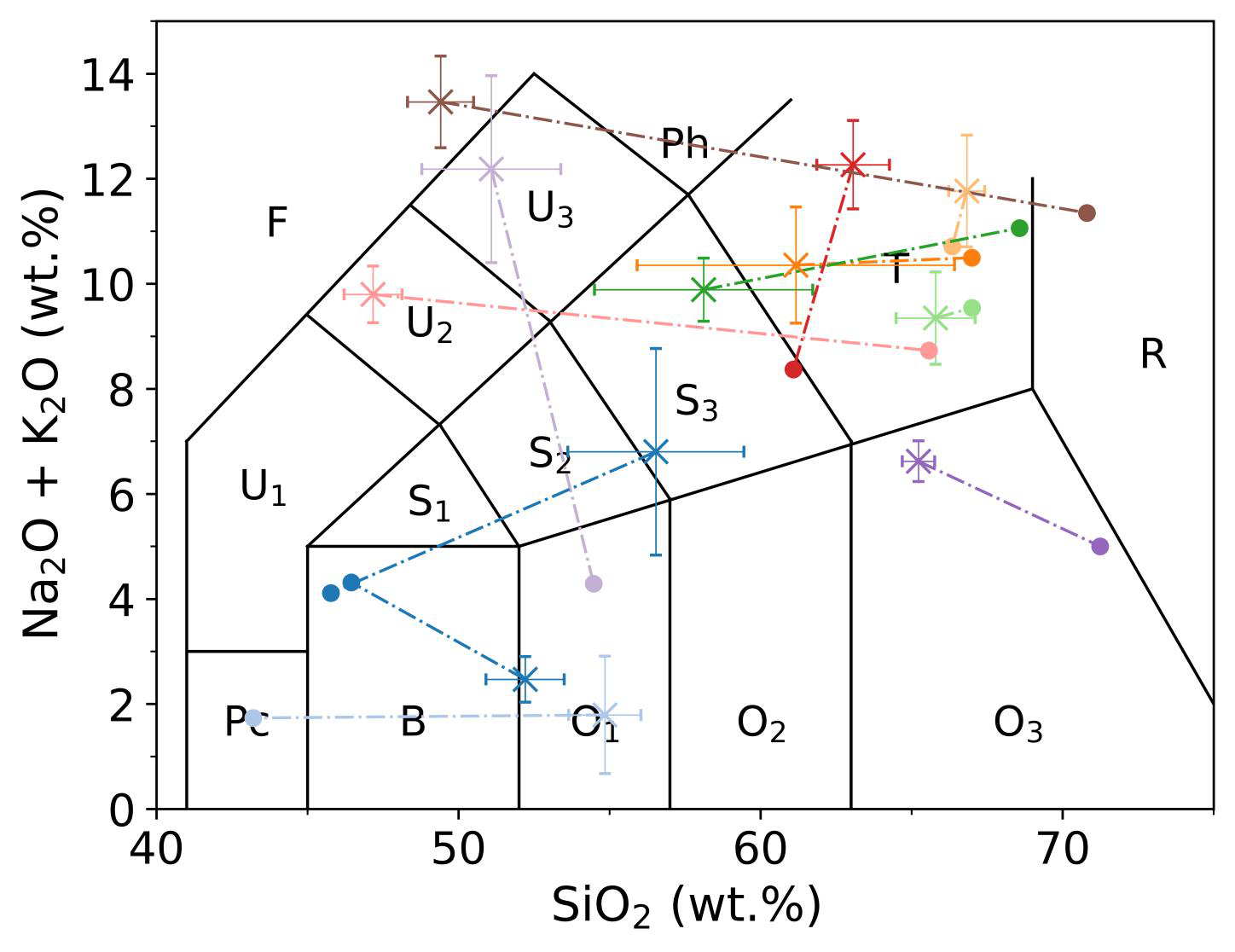}
\caption{TAS classification using machine learning models of the validation rocks. The positions determined by XRF are presented in colored solid circles, the corresponding model-predicted positions are presented in the same color in crosses. A dashed line links the XRF reference position and the model-predicted position of a same rock sample. The error bars on the predicted position are calculated among the different pretreated spectra of a given validation sample.}
\label{fig:fig7}
\end{figure}

\subsection{Analytical Performances with the transfer learning model.}

In parallel to the results with the machine learning models and in order to review the improvements with the transfer learning models, the calibrations models resulted from transfer learning are shown in Figure 8. The extracted parameters for assessment of the model performances are presented in Table 2. In Table 2, we can see that although the transfer learning models present slightly lower calibration performance in terms of $r^2$, the slope, $LOD$, $REC$ and $RET$, the performance for prediction for rock samples are significantly improved, especially for $REP$. This means that the participation of the 8 rock samples in the training data set together with the retained pellet samples efficiently takes into account the physical matrix effect and reinforces the robustness of the models for prediction with rock samples. Correspondingly in Figure 8, we can see significant reductions of bias and variance of the predicted concentrations for the validation rock samples with respect to the calibration models trained with a part of the pellet samples and the training rock samples. In particular, for $SiO_2$, 18 pellet samples were retained in the training data set among the 20 ones by the optimization loop during the model training process. The retained pellet sample were respectively 13 and 14 for $Na_2O$ and $K_2O$. 

The calibration models shown in Figure 8 were used to represent the validation rock samples in a TAS diagram. The obtained result is shown in Figure 9 using the same symbols as in Figure 7. We can see a much improved result conforming the good performance of the transfer learning models shown in Figure 8 and Table 2. A detailed counting shows 10 correctly classified validation rock samples. Only two samples were classified into a wrong field (S4 and S6), although they are very close to borders separating the correct and wrong fields. The rate of correct classification can thus be calculated to be 88.3\%. 

\begin{figure}
\centering\includegraphics[width=7cm]{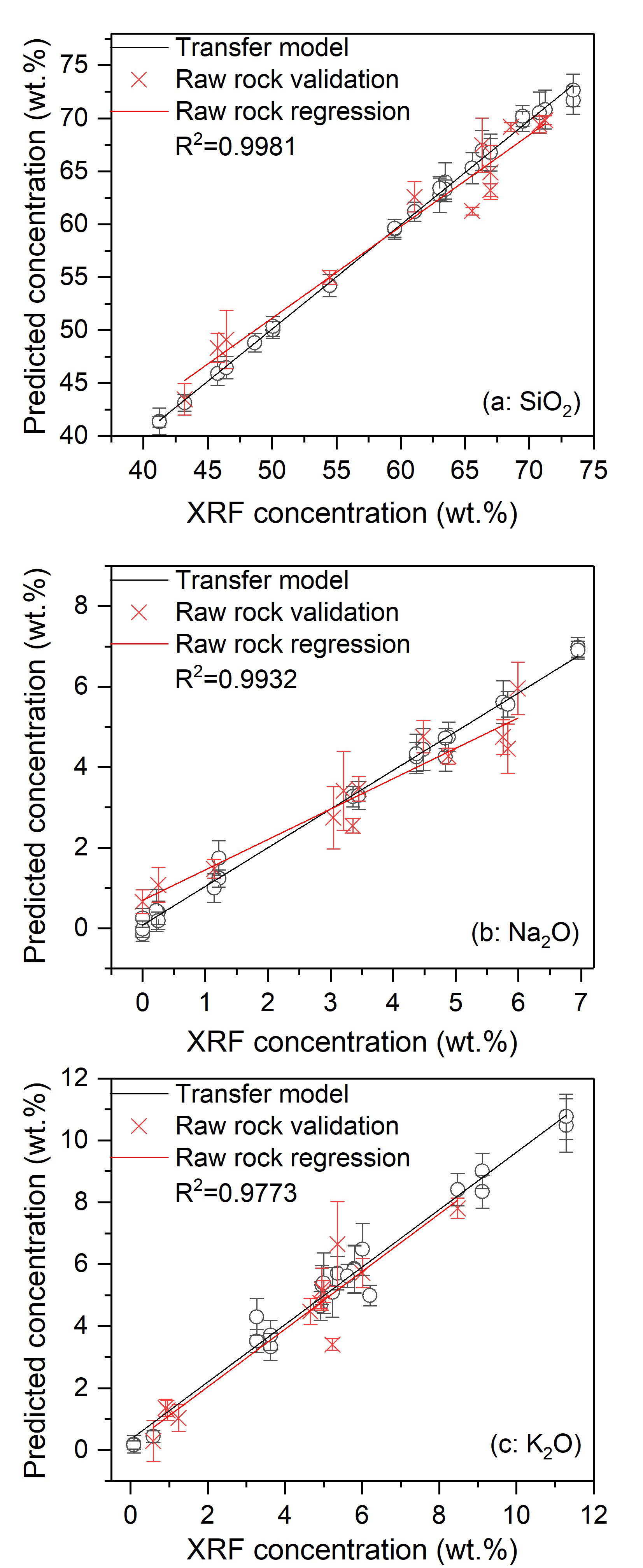}
\caption{Transfer learning-based calibration models trained with the pellet samples and the training set of the rock samples (black lines) together with the calibration data (black open cycles) for the 3 compounds $SiO_2$ (a), $Na_2O$ (b) and $K_2O$ (c). Validation data from the rock validation samples are presented in red crosses, their linear regressions in red lines. The error bars of the presented data correspond to the standard deviations ($\pm SD$) of the predicted concentrations over the 21 pretreated spectra for a given sample.}
\label{fig:fig8}
\end{figure}

\begin{table}
  \caption{Parameters assessing the calibration and prediction performances of the transfer learning calibration models for $SiO_2$, $Na_2O$ and $K_2O$.}
  \label{tbl:2}
  \begin{tabular}{llllll}
    \hline
    \multicolumn{2}{c}{Compound} & $SiO_2$ & $Na_2O$ & $K_2O$ & Average\\
    \hline
    \multirow{5}{*}{Calibration} & $r^2$ & 0.998 & 0.993 & 0.997 & 0.996\\
    & $Slope$ & 0.985 & 0.961 & 0.993 & 0.980\\
    & $LOD(\%)$ & 3.70& 1.03 & 2.26 & 2.33\\
    & $REC(\%)$ & 5.61 & 15.7 & 6.40 & 9.25\\
    & $RET(\%)$ & 4.90 & 16.2 & 7.72 & 9.59\\
    \hline
        \multirow{2}{*}{Validation} & $REP(\%)$ & 3.11 & 41.99 & 19.79 & 21.63\\
    & $RSD(\%)$ & 24.07 & 21.29 & 101.1 & 48.82\\
    \hline
  \end{tabular}
\end{table}

\begin{figure}
\centering\includegraphics[width=7cm]{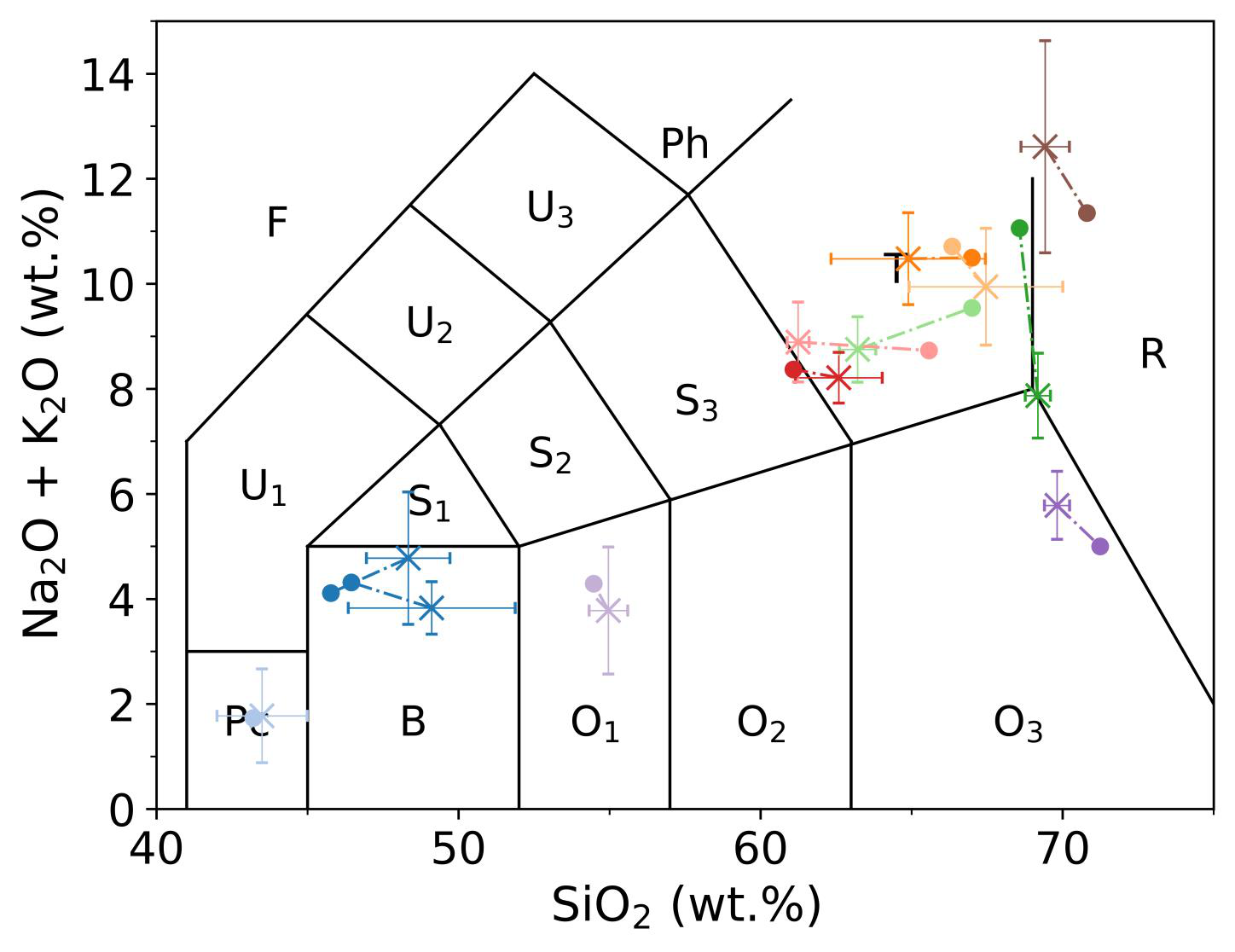}
\caption{ TAS classification using transfer learning models of the validation rocks. The used symbols are similar to those used in Figure 7.}
\label{fig:fig9}
\end{figure}

\section{CONCLUSIONS}

In this work, within a specific application of classification of rocks using the TAS diagram, we have introduced transfer learning in LIBS spectral data treatment to improve the performance of the models trained with laboratory standard samples in the form of pellets when used for prediction with LIBS spectra from natural rock samples. Such scenario corresponds well to many important applications, such as Mars exploration with LIBS, where LIBS spectra acquired \textit{in situ} using LIBS instruments onboard Mars rovers are treated with prediction models established using laboratory-prepared standard samples. Obvious differences in physical state between the laboratory standards and the natural rocks analyzed on Mars lead to unavoidable physical matrix effect that needs to be corrected for accurate and precise determination of compound concentrations, which is the basis of the TAS classification of Martian rocks, among other geochemistry analyses, very important for the scientific objectives fixed for the Mars exploration missions. 

In particular, feature-representation-transfer and instance-transfer as the two important features of transfer learning were implemented in the LIBS spectral data treatment. A set of common features was thus determined jointly by the ensembles of features respectively selected for the pellet samples and the training rock samples. These common selected features were thus used as the input variables for the training and validation of the transfer learning models. Instance-transfer consisted in retaining among the pellet samples, those who are efficient to improve the performance of the trained model in a procedure of cross validation with the training rock samples. The performances of the transfer learning models were compared with those of the machine learning models. Significant improvements have been observed for predictions with the LIBS spectra from the rock validation samples for the 3 concerned compounds involved in the TAS classification, $SiO_2$, $Na_2O$ and $K_2O$. The rate of correct TAS classification has been improved from 33.3\% with the machine learning models to 88.3\% with the transfer learning models.

Our work therefore demonstrates the efficiency of transfer learning in the treatment of rock LIBS spectra using machine learning-based models, once a suitable set of rock samples are included in the model training process. Beyond Mars exploration with LIBS, similar scenarios exist also in LIBS industrial applications. Our findings in this work can have thus a more general interest in the development of LIBS technique for various applications.

\subsection{Author Contributions}

CS studied and developed the data treatment method, wrote the corresponding computer programs, and wrote the draft of the paper. WX prepared the samples and acquired the LIBS spectra. YT participated in the development of the feature selection algorithm. YZ, ZY, SS, MW, LZ, FC participated in the experimental setup development and LIBS spectrum acquisition. JY supervised the research program and wrote the paper. The manuscript was written through the contributions of all the authors. All authors have given the approval to the final version of the manuscript.

\subsection{Notes}
The authors declare no competing financial interest.

\begin{acknowledgement}

This work was supported by the Startup Fund for Youngman Research at SJTU, the National Natural Science Foundation of China [Grants 11574209, 11805126, 61975190].

\end{acknowledgement}


\bibliography{achemso-demo}

\end{document}